# Second-harmonic generation in periodically-poled thin film lithium niobate wafer-bonded on silicon

ASHUTOSH RAO,[1] MARCIN MALINOWSKI,[1] AMIRMAHDI HONARDOOST,[2] JAVED ROUF TALUKDER,[1] PAYAM RABIEI,[3] PETER DELFYETT,[1] AND SASAN FATHPOUR[1,2,*]

[1]*CREOL, The College of Optics and Photonics, University of Central Florida, Orlando, FL, 32816, USA*
[2]*Department of Electrical and Computer Engineering, University of Central Florida, Orlando, FL 32816, USA*
[3]*Partow Technologies LLC, Orlando, FL 32816, USA*
[*]*fathpour@creol.ucf.edu*

**Abstract:** Second-order optical nonlinear effects (second-harmonic and sum-frequency generation) are demonstrated in the telecommunication band by periodic poling of thin films of lithium niobate wafer-bonded on silicon substrates and rib-loaded with silicon nitride channels to attain ridge waveguide with cross-sections of ~ 2 $\mu m^2$. The compactness of the waveguides results in efficient second-order nonlinear devices. A nonlinear conversion of 8% is obtained with a pulsed input in 4 mm long waveguides. The choice of silicon substrate makes the platform potentially compatible with silicon photonics, and therefore may pave the path towards on-chip nonlinear and quantum-optic applications.

## References and Links

## 1. Introduction

Quasi-phase matching (QPM) for three-wave mixing has enabled progress in fields such as telecommunications [1] and quantum optics [2,3]. Numerous three-wave optical mixers on-chip have been investigated towards high performance integration with material systems such as silicon and III-V compounds. Silicon (Si) lacks intrinsic second-order nonlinearity ($\chi^{(2)}$), essential for efficient nonlinear three-wave frequency mixing, due to its centrosymmetric crystalline nature. Nonlinear effects demonstrated on the silicon-on-insulator (SOI) platform are thus typically based on third-order nonlinearity ($\chi^{(3)}$), which is significantly weaker than nonlinear mixing driven by the $\chi^{(2)}$ tensor. Furthermore, important three-wave mixing nonlinear effects such as second-harmonic generation (SHG) are conveniently realized using $\chi^{(2)}$ nonlinearity, and are much more difficult to implement using $\chi^{(3)}$ nonlinearity. SHG has been pursued in III-V waveguides [4,5]. However, these attempts are limited by high optical propagation losses and difficulties of poling.

It is well-known that lithium niobate (LiNbO$_3$ or LN) possesses one of the highest $\chi^{(2)}$ values and a broad transmission window [6]. Z-cut bulk periodically-poled lithium niobate (PPLN) waveguides have thus been established as the method of choice for implementing QPM for efficient three-wave mixing [7]. The in-diffusion of titanium into bulk LN wafers [8] and annealed proton exchange [9] are two methods used to define conventional LN stripe waveguides for QPM. However, these conventional diffused LN waveguides suffer from low index contrast and typically work only for transverse-magnetic (TM) waveguide modes. This leads to poor optical confinement, large mode sizes, and poor overlap between different optical modes for three-wave mixing, such as the fundamental and second-harmonic modes involved in SHG and sum-frequency generation (SFG). PPLN waveguides have been reported in the past on bulk Z-cut [10] and X-cut [11-13] LN with normalized nonlinear conversion efficiencies around 40 %W$^{-1}$cm$^{-2}$ [10-12] in telecom wavelengths. One approach to integrate LN on Si is metallic bonding using gold, where ~ 10 µm thick films of LN are bonded after poling on Si, with conversion efficiencies around 80 %W$^{-1}$cm$^{-2}$ [14]. Recently, there has been an interest in poling X-cut thin film LN [15,16]. One effort based on thin film X-cut PPLN on a LN substrate has demonstrated a nonlinear conversion efficiency around 160 %W$^{-1}$cm$^{-2}$ [16].

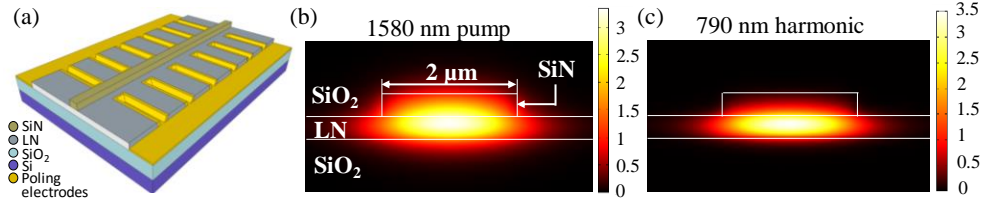

Fig. 1. (a) Schematic of the device depicting the silicon nitride (SiN) rib, the lithium niobate (LN) slab, the silicon dioxide (SiO$_2$) lower cladding, the silicon (Si) substrate, and the metal poling electrodes. The SiO$_2$ top cladding is excluded for clarity; (b) & (c) COMSOL$^{TM}$ simulations of the fundamental TE waveguide modes at the pump wavelength (1580 nm) and the second harmonic wavelength (790 nm).

Evidently, the most widely used QPM devices based on bulk PPLN are not directly compatible with state-of-the-art integrated photonics on silicon substrates or suffer from low efficiency due to large optical mode size. Hybrid submicron waveguides of LN integrated onto silicon substrates would be the ideal solution for introducing efficient $\chi^{(2)}$ photonics on silicon. We have previously solved the challenge of obtaining tightly confined LN waveguides by ion implantation and slicing of thin films of LN heterogeneously integrated onto an oxidized silicon substrate, and rib loading the films with index-matching materials [17,18]. Different materials, such as tantalum pentoxide and chalcogenide glass have been previously developed for low loss [19-21] and used for rib loading the house-made LN thin films for demonstrating optical modulators [17,18]. Another index-matching alternative material with a wide transmission window used by us for modulators is silicon nitride (SiN) [22].

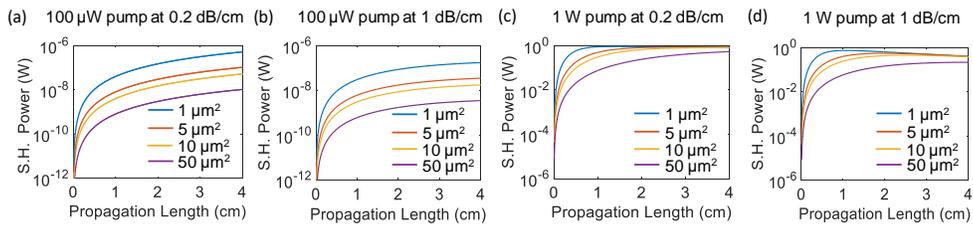

Fig. 2. Numerical simulations for the generation of second harmonic (S.H.) power using PPLN waveguides across different input CW pump powers and propagation losses for varying effective mode overlap areas: (a) and (b) For an input power of 100 µW, there is about an order of magnitude improvement in nonlinear conversion for submicron waveguides, even for propagation lengths up to 4 cm, irrespective of propagation loss; (c) and (d) For an power of 1 W, the nonlinear conversion offered by submicron PPLN waveguides is ~ 50% in less than 1 cm of propagation, even at a relatively high loss of 1 dB/cm, while conventional PPLN solutions require much longer lengths.

A schematic of the PPLN on Si device with SiN ribs presented in this work is shown in Fig. 1(a). Using this structure, the first integration of a locally-poled PPLN wavelength converter on a silicon substrate is demonstrated. Unlike previous PPLN on Si efforts [14], the poling in this work is performed locally after wafer bonding, offering greater flexibility than directly bonding pre-poled LN on Si. Besides, the present ion-slicing approach [17,18] avoids metallic bonding and the high optical loss associated with it.

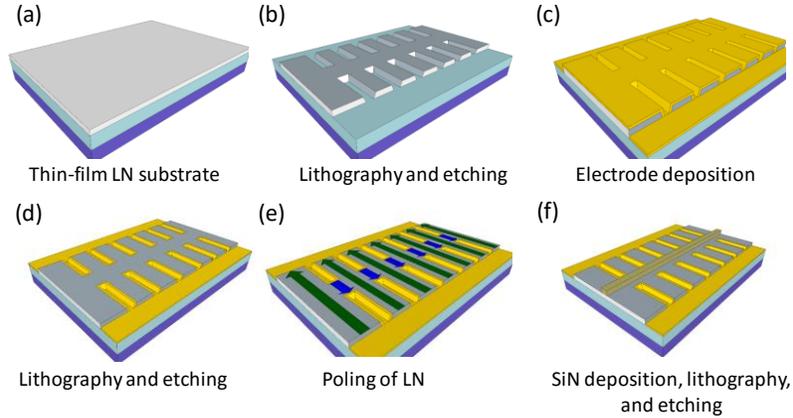

Fig. 3. Major fabrication steps (a) Y-cut LN on Si substrate (b) First lithography and etching of LN; (c) Metal electrode deposition; (d) Lithography and etching to completely define the periodic electrodes; (e) Poling of LN on Si with periodic domain reversal; (f) SiN rib definition by PECVD, lithography, and etching to form the ridge waveguide. Not shown in this figure is the final deposition of a $SiO_2$ top cladding by PECVD.

## 2. Design

The slab region of the ridge optical waveguide is formed by bonding a 400-nm-thick film of Y-cut LN to a lower cladding 2-μm-thick layer of SiO2 on a Si substrate. The crystal orientation of the LN thin film is chosen to utilize the highest nonlinear coefficient of LN, viz., $d_{33}$ = 30 pm/V [6], for the transverse-electric (TE) waveguide modes. The z-axis of the LN thin film is thus aligned along the electric field of the TE mode. Waveguides are formed by rib-loading the LN thin films with strips or channels of SiN (Fig. 1(a)). COMSOL$^{TM}$ simulations are used to determine the poling periodicity required for the QPM of the SHG process with a pump wavelength in the telecommunication band, based on the dimensions of the SiN strips. The SiN ribs are 400 nm tall and 2,000 nm wide, resulting in a poling period around 5 μm for TE polarized pump and harmonic light. Around 65% of the pump and 90% of the second harmonic fundamental TE optical modes are confined in the LN thin film. The modes, simulated with COMSOL$^{TM}$, are shown in Figs. 1(b) and 1(c). The simulated value of the normalized conversion efficiency is around 1400 %$W^{-1}cm^{-2}$ for including propagation loss, with a mode overlap integral around 2.3 $\mu m^2$ at a pump wavelength of 1,580 nm. Additional electric field simulations run in COMSOL$^{TM}$ are used to determine the duty cycle of the poling electrodes. The high optical confinement leads to significantly enhanced nonlinear interaction, thereby reducing the device length required for nonlinear conversion. This is verified in Fig. 2, where the generated second harmonic power is numerically simulated across a wide range of effective mode overlap areas for different input continuous-wave (CW) power levels and propagation losses. The figure indicates a clear increase in nonlinear power conversion with increasing modal confinement for shorter propagation lengths, irrespective of the propagation loss.

## 3. Fabrication

Similar to our prior works on optical modulators [17,18], ion implantation and room-temperature bonding are used in-house to transfer *Y*-cut LN thin films onto thermal $SiO_2$ cladding layers on silicon substrates in accordance with the design and dimensions described above. The fabrication steps are depicted in Figure 3. Electron-beam lithography on a Leica EBPG5000+ writer, followed by dry etching using inductively-coupled plasma reactive-ion etching (ICP-RIE), was used to define the boundaries of the periodic electrodes. The dry etch was engineered to produce a pronounced LN side wall angle around 70°. This ensured good contact between the subsequent metal deposition and the LN sidewall for the entire depth of the LN film. The electrodes are sufficiently far away (> 3 μm) from the waveguide code (SiN rib) to avoid metallic loss of the optical waves. Metal poling electrodes with a 30 % duty cycle and 9 μm separation were fabricated using electron beam evaporation of 100 nm of chromium, lithography, and dry etching. The dies were poled across the etched ridges by applying high voltage pulses to the poling electrodes using contact probes at room temperature. Next, the SiN rib layer was deposited using plasma enhanced chemical vapor deposition (PECVD). This was patterned using electron-beam lithography, and dry-etched to form the rib-loaded region. Finally, a 2-μm-thick $SiO_2$ top cladding was deposited using PECVD, and the dies were diced and polished.

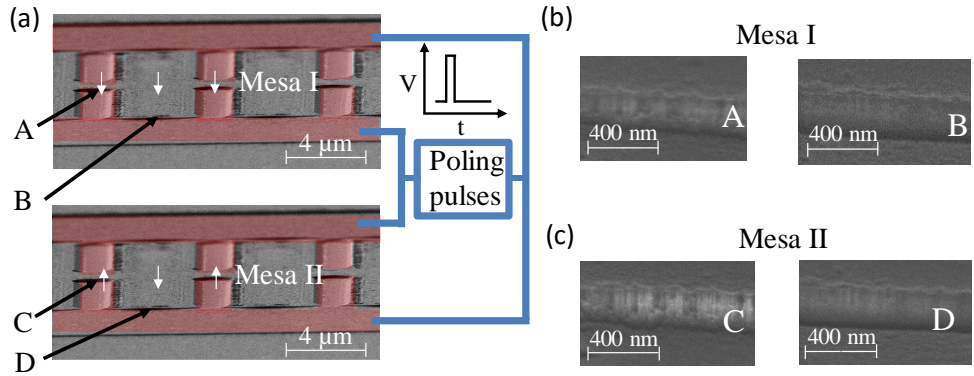

Fig. 4. (a) False color SEMs of an unpoled and poled LN mesa with a schematic depiction of the high voltage pulses applied with opposing polarity across each mesa; (b) & (c) Surfaces A-D after the high voltage poling pulse is applied to both LN mesas simultaneously and etching in hydrofluoric acid. Surface C is significantly etched compared to surfaces A, B and D as seen in the SEMs, indicating successful domain reversal.

During poling, the electric field was ramped up to ~ 40 kV/mm, with voltages around 350 to 400 V, for 10 ms and ramped down slowly to prevent back-switching of the inverted domains [23]. The high electric field was used to ensure the onset of nucleation and subsequent domain reversal. Domain merging was avoided by using three identical pulses and limiting the pulse duration to 10 ms [24]. The domains are expected to be poled entirely through the LN film based on the high poling field, the long pulse duration, and the deep LN etching for the poling electrodes. To confirm the periodic reversal of the lithium niobate domains, the high voltage poling pulses were applied simultaneously across two separate pairs of electrodes across mesas on the same die, but with opposite polarity. Thus, only one of the mesas would be poled. This scheme is depicted in Fig. 4(a). After poling, the metal electrodes were wet-etched away, and the die was briefly etched in hydrofluoric acid. The sidewalls of the two LN mesas, labelled A-D in Fig. 4(a), were imaged using a scanning electron microscope (SEM) (Figs. 4(b) and (c)). Evidently, Surfaces B and D were etched the least, as they were not exposed to the high poling field and thus could not have undergone domain reversal. Surfaces A and C were both exposed to the high poling fields, and were thus etched more than B and D. This is attributed to the electric field right under the poling electrodes, typically being up to an order of magnitude higher than the breakdown strength of bulk dielectrics [23]. Finally, sidewall C was etched the most, clearly indicated in Fig. 4(c). This was in keeping with the differential etching of the positive and negative z-axis of LN [25], and confirms successful poling.

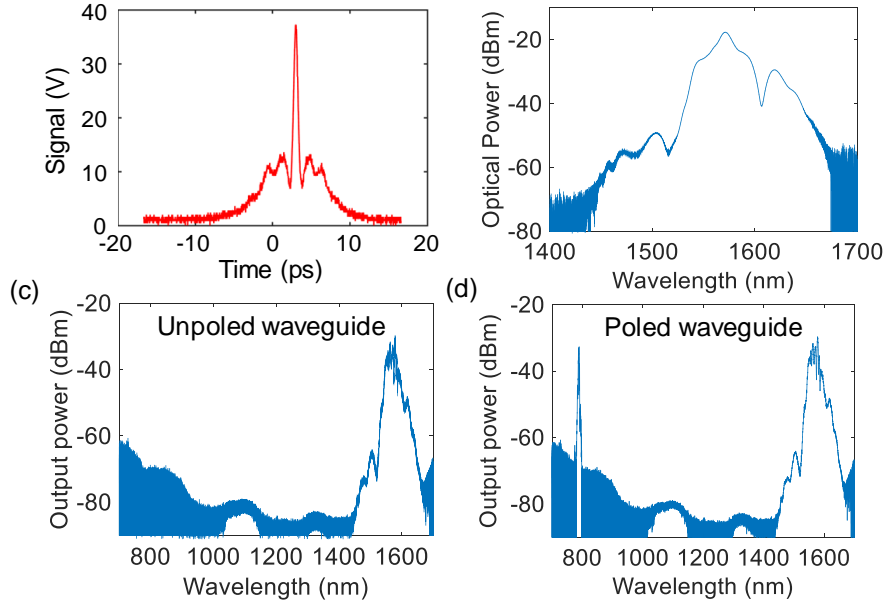

Fig. 5. (a) Autocorrelation and (b) optical spectrum, of the pulsed input to the PPLN waveguide; (c) Output of a reference unpoled waveguide; (d) Output of a poled waveguide, with a frequency doubled signal around 788 nm.

## 4. Characterization

Obviously, the ultimate proof for successful poling is demonstration of nonlinear optical processes that require QPM. The fabricated 4-mm long thin film PPLN waveguides were optically characterized by coupling light in and out of the chips by lensed fibers. The input was a mode-locked fiber laser operating at a MHz repetition rate amplified by an erbium-doped fiber amplifier, followed by a polarization controller. The propagation losses at the pump and second harmonic wavelength are < 1 dB/cm, obtained by analyzing the insertion loss across a number of waveguides. The material absorption of the SiN rib material is < 1 dB/cm and < 0.2 dB/cm at the fundamental and second harmonic wavelength respectively, measured on a Metricon prism-coupling system. The coupling loss measured per facet at the pump wavelength is ~ 10 dB. The coupling loss at the second harmonic wavelength is marginally higher due to the multimode nature of the lensed fibers used. The coupling losses can be potentially reduced by the use of appropriately-designed input and output waveguide tapers, or the use of free-space lensed input and output coupling.

Figure 5(a) shows the autocorrelation of the input pulse. The input pulse is around 500 fs wide, sitting on a 7 ps wide pedestal. While the pedestal lowers the peak power of the input and generated output pulse, it is sufficiently wide to diminish the effects of pulse walk-off induced by group-velocity mismatch induced in the 4 mm long devices [26-29]. The optical spectrum of the input pulse is shown in Fig. 5(b). The output after the PPLN samples was fed directly to an optical spectrum analyzer. An optical power meter was used alternatively to monitor the output power after the waveguide. The optical spectrum recorded for the reference unpoled waveguide is shown in Fig. 5(c), with no evidence of harmonic generation. The poled waveguide phase-matches near a fundamental wavelength of 1,580 nm. The output optical spectrum is shown in Fig. 5(d), with a clear frequency doubled signal generated around 788 nm. The average second-order nonlinear conversion efficiency of the device is extracted to be 8% by integrating the input and output average power. This effective efficiency includes contributions from any phase-matched second-order process, i.e., primarily SHG, although SFG could have considerable contribution too. It is practically difficult to differentiate the weight of each effect, due to the pulsed nature of the input source.

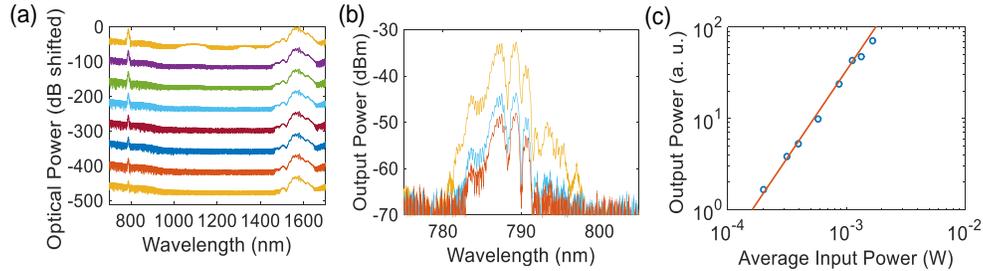

Fig. 6. (a) Optical spectra for decreasing input pump powers (top to bottom), displaced by 60 dB. The input power for each trace can be read in part (c); (b) Optical spectra around the generated output signal wavelength of 3 of traces in part (a) with input average powers of 1.67, 0.87 and 0.31 mW, respectively; (c) A straight line fit of slope 1.91 on a logarithmic scale shows the quadratic dependence of the output signal on the input pump.

Optical spectra, displaced by 60 dB, recorded at decreasing input pump powers, are plotted in Fig. 6(a). Figure 6(b) shows the generated output signals of the 1st, 4th, and 7th traces, from the top, plotted in Fig. 6(a). The slight asymmetry of the spectrum in Fig. 6(b) could be attributed to the asymmetry of the contribution of SHG to the generated signal. The quadratic nature of the generated signal is seen in Fig. 6(c), where the average integrated output power is plotted against the average integrated input power on a logarithmic scale from the traces in Fig. 6(a). A linear fit to these measurements yields a slope of 1.91, which is very close to the slope of 2 expected for low propagation loss.

## 5. Conclusions

The first compact heterogeneous periodically-poled thin film LN waveguides on silicon substrates have been fabricated and characterized. A nonlinear conversion of 8% has been obtained with a pulsed input. These devices are 4 mm long and the cross-sections are significantly smaller than traditional PPLN wavelength converters. The compact size along with the use of a silicon substrate demonstrates the compatibility of efficient $\chi^{(2)}$-based nonlinear photonic devices with silicon photonics for potential on-chip nonlinear and quantum-optic applications.

## Funding

This project is being supported by the U.S. Office of Naval Research (ONR) Young Investigator Program and the U.S. Defense Advanced Research Projects Agency (DARPA). The views, opinions and/or findings expressed are those of the author and should not be interpreted as representing the official views or policies of the Department of Defense or the U.S. Government.